\documentclass[prl,twocolumn,amsmath,amssymb,showpacs,superscriptaddress,floatfix]{revtex4}

\usepackage{graphicx}
\usepackage{color}
\usepackage{soul}
\usepackage{epsfig}
\usepackage{float}

\begin{document}
\title{Orientational Order in Disordered Colloidal Suspensions}

\date{\today}

\author{Xiunan Yang$^*$}
\affiliation{Beijing National Laboratory for Condensed Matter Physics and Key Laboratory of Soft Matter Physics, Institute of Physics, Chinese Academy of Sciences, Beijing 100190, People's Republic of China}
\affiliation{University of Chinese Academy of Science, Beijing 100049, People's Republic of China}
\author{Wei-Hua Wang$^*$}
\affiliation{Institute of Physics, Chinese Academy of Sciences, Beijing 100190, People's Republic of China}
\affiliation{University of Chinese Academy of Science, Beijing 100049, People's Republic of China}
\author{Ke Chen$^*$}
\affiliation{Beijing National Laboratory for Condensed Matter Physics and Key Laboratory of Soft Matter Physics, Institute of Physics, Chinese Academy of Sciences, Beijing 100190, People's Republic of China}
\affiliation{University of Chinese Academy of Science, Beijing 100049, People's Republic of China}

\begin{abstract}
Exploring structural order in disordered systems including liquids and glasses is an intriguing but challenging issue in condensed matter physics. Here we construct a new parameter based on the angular distribution function of particles and show that this new orientational order has significantly higher correlation with dynamic heterogeneity compared to a translational parameter based on the radial distribution functions in colloidal glasses. The gradual development of orientational and translational order in supercooled liquids shows that the higher correlation between orientational order and dynamics comes from the onset of glass transition and the orientational order would dominate the glassy dynamics after a simple liquid is considerably supercooled. Our results suggest that orientational order reflects the formation of amorphous order during glass transition while translational order is mainly a result of density increase. 
\end{abstract}

\maketitle
Glassy systems such as supercooled liquids and amorphous solids are refferred to as "disordered" matters, due to the lack of long-range translational order. More puzzling is that there are slight structural changes at the level of radial distribution functions (RDFs) when glasses form from liquids with dynamics slowing down dramatically~\cite{Grfail1,Grfail2}. However, investigations on glassy dynamics suggest that glasses and supercooled liquids are not fully random. Non-Gaussian heterogeneous dynamics with non-exponential decay are observed in a wide variety of glassy systems~\cite{DH1,DH2,DH3,DH4,DH5,DH6,DH7,DH8,DH9,DH10}. An increase of dynamical correlation length with packing fraction increasing or temperature declining suggests that there are correlations or some sort of order under the appparent disorder~\cite{DH11}. Studies on sophisticated structural order such as symmetry-based models~\cite{Tanaka1,sym_2,sym_3,sym_4,sym_5} and point-to-set correlations~\cite{PTS1,PTS2,PTS3,PTS4,PTS5} also suggest that the dynamic heterogeneity is a consequence of structure order. Nevertheless, detailed identifications of amorphous structure and a casual link between structure and glassy dynamics remain elusive~\cite{Paddy1}. 

Recently, machine-learnt softness~\cite{Andrea1,Andrea2,Andrea3,Andrea4} and particle-level structural entropy $S_{2}$~\cite{Yang2,Tanaka1,Tanaka2,Han1} embody promises of revealing structural orders in disordered systems for binary glassy systems with strong geometrical frustrations. The efficiency of these two structural parameters, which directly use or can be approximated by using the local RDFs~\cite{Andrea1, Yang2}, suggests that the configuration information in local RDF are essential to characterize amorphous order. As a whole, the distribution function of amorphous systems are assumed isotropic and thus ensemble-averaged RDFs are usually angle-averaged. For a single particle, however, the several shells of particles surrounding it are not uniformly distributed by angle. Therefore, the local angular distribution function (ADF) could not be neglected. Besides, ADF is a physically appealing way of considering many-body effects ~\cite{Tanaka3}, which are ignored by two-body RDFs or their derivatives.

\begin{figure*}

    \includegraphics[width=16cm]{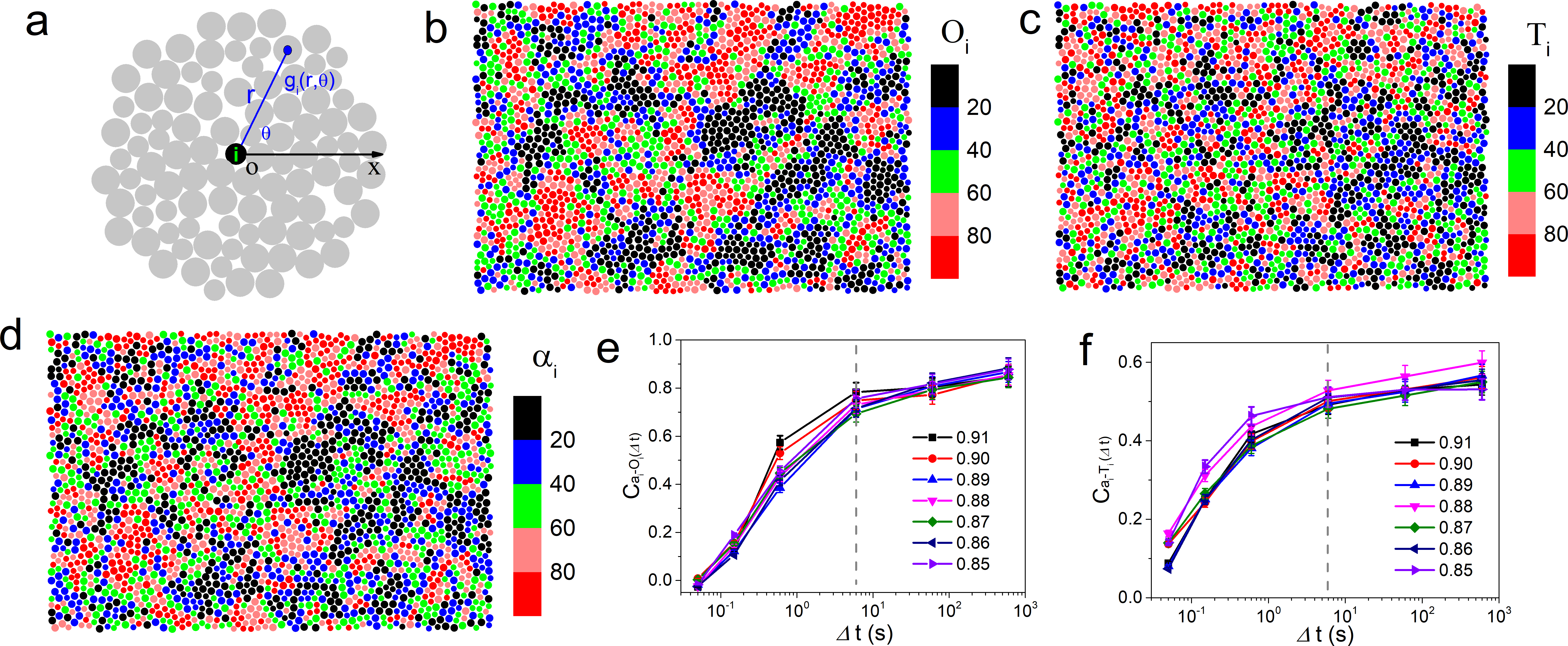}
  \caption{\textcolor[rgb]{0.98,0.00,0.00}{(color online)}
\textbf{ Order parameters in colloidal glasses.} {\bf a,} Particles in the polar coordinate system with reference particle i as pole O and with x-axis as polar axis. We use $g_{i}(r,\theta)$ to characterize the probability of finding particles at the point $(r,\theta)$ with radial coordinate $r$ and angular coordinate $\theta$. Spatial distribution of {\bf b,} orientational fluctuation function $O_{i}$, {\bf c,} translational fluctuation function $T_{i}$, and {\bf d,} local Debye-Waller factor $\alpha_i$. All three parameters are plotted by the ranks of individual particles. Spearman's rank correlation between $\alpha_i$ and {\bf e,} $O_{i}$, {\bf f,} between $\alpha_i$ and $T_{i}$.}
  \label{fig1}
\end{figure*}

In this letter, we construct translational and orientational structural orders based on local RDF and local ADF. These parameters characterize the broken symmetry in glasses and supercooled liquids compared to ideal gas. By comparison, we find that the orientational order plays dominant roles in dynamic heterogeneity of colloidal glasses. In contrast to the slight change of translational order in liquids with packing fraction increasing consistently, the orientational order grow dramatically and are mostly responsible for dynamic heterogeneity after liquids are supercooled. The maximally preferred directions are also shown to be correlated within three shells of neighbours.

We used colloidal particles as big 'atoms'. A binary mixture of poly-N-isopropylacrylamide (PNIPAM) particles between two cover slips was hermetically sealed to prepare two-dimensional colloidal suspensions~\cite{Yunker1, Chen2010, Chen2011}. The binary mixtures with number ratio 1:1 and with diameter ratio 1:1.4 were used to avoid crystallization. The particles are thermo-sensitive and the temperature could be controlled by thermal coupling to the microscope objective (BiOptechs). Thus we could tune the samples' packing fraction in-situ from high-packing amorphous solids to low-packing liquids. There were two groups of packing variations consisting of $\sim3500$ particles. One group with packing fraction varying from 0.91 to 0.85 was used to simulate the glassy solids; the other group was a series of liquids with packing fractions between 0.54 and 0.83. Before all the data acquisitions, we relaxed the samples for more than $3$ hours. After each temperature change, the samples were equilibrated for more than $15$ minutes before another measurement. The particle configurations were recorded by standard video microscopy at $60-110$ frames/s,and particle trajectories were extracted by particle-tracking techniques~\cite{Grier1}.

In glassy systems, each particle is surrounded by dozens of other particles as in Fig.~\ref{fig1}a. A polar coordinate system is used to describe the local distribution function for particle $i$ $g_{i}(r,\theta)$. $g_{i}(r,\theta)$ is defined as the probability of finding particles at radial coordinate $r$ and angular coordinate $\theta$. Normalized by the uniform distribution in ideal gas, $g_{i}(r,\theta)=n(r,\theta)/n_{e}(r,\theta)$. $n_{e}(r,\theta) =constant *r$ is the expected number of particles in a small ring with radius $r$ in two-dimensional ideal gas. By an angular average, we obtain local RDF $g_{i}(r) =\sum_{\theta} g_{i}(r,\theta)$; with a radial average, local ADF is $g_{i}(\theta) =\sum_{r} g_{i}(r,\theta)$.
For comparisons between local RDFs or local ADFs of different particles, we normalize $g_{i}(\theta)$ as $p_{i}(\theta)=\frac{g_{i}(\theta)}{\sum_{\theta} g_{i}(\theta)}$; $g_{i}(r)$ as $p_{i}(r)=\frac{g_{i}(r)}{\sum_{r} g_{i}(r)}$.

Structure emerges when symmetries break down and the distribution functions become nonuniform. Based on this philosophy, we use the fluctuations of local RDFs and local ADFs to characterize the structure of each particle. There are two algorithms to quantify that fluctuations. An intuitive choice is the variance of $p_{i}(\theta)$ or $p_{i}(r)$. In our binary systems, the orientational and translational fluctuations are defined as $O_{i} =-\sum_{u=b,s} \rho_{i,u}\sigma_{i,u}(\theta)$ and $T_{i}=-\sum_{u=b,s} \rho_{i,u}\sigma_{i,u}(r)$, where $\rho_{i,u}$ are the number density of big (b) or small (s) particles surrounding particle i, $\sigma_{i,u}(\theta)~(or~\sigma_{i,u}(r))$  are the variance of $p_{i,u}(\theta)~(or~p_{i,u}(r))$.
The other algorithm is the nominal entropy of $p_{i,u}(\theta)~(or~p_{i,u}(r))$: $S_{i}^O=-\sum_{u=b,s}\sum_{\theta} \rho_{i,u}p_{i,u}(\theta)\ln p_{i,u}(\theta)$ and $S_{i}^T=-\sum_{u=b,s}\sum_{r} \rho_{i,u}p_{i,u}(r)\ln p_{i,u}(r)$. The correlation between $O_{i} (T_{i})$ and $S_{i}^O(S_{i}^T)$ is as high as 0.99, indicating the equality between $O_{i} (T_{i})$ and $S_{i}^O(S_{i}^T)$ as good metrics of local RDF (or local ADF) fluctuations. In this Letter, we use the variances $O_{i}$ and $T_{i}$ to characterize the fluctuations. The fluctuations are calculated from three shells of neighbors surround each particle to avoid the loss of a large fraction of particles due to the boundary effect. More shells of neighbors contribute trivially to the correlation between the order parameters and dynamics as in supplementary~\cite{suppl.} since particles far from the reference particle are more uniformly distributed.

\begin{figure*}
    \includegraphics[width=16cm]{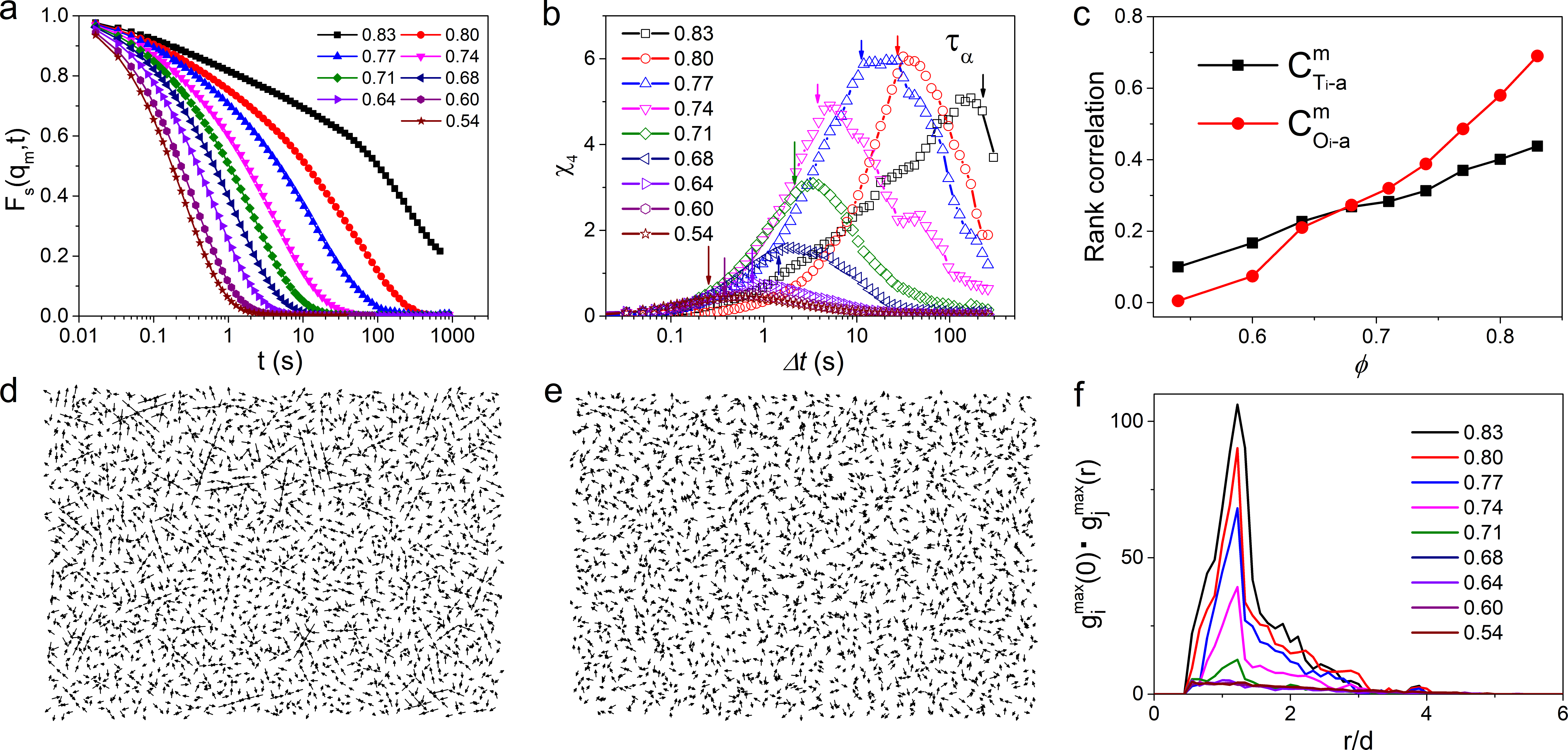}
  \caption{\textcolor[rgb]{0.98,0.00,0.00}{(color online)} \textbf{ Emergence of orientational order during glass transition.} {\bf a,} Self intermediate scattering functions for liquids in different packing fractions $\phi$. {\bf b,} The dependence of $\chi_4$ on duration $\Delta t$. $\chi_4$ reaches its maximum value with $\Delta t$ close to $\alpha$-relaxation time as indicated by the arrows. {\bf c,} $\phi$ dependence of the rank correlation between $\alpha_i$ and distribution fluctuations $O_{i}$ ($T_{i}$). There is a crossover at $\phi_c\sim0.7$ where orientational order becomes better correlated with dynamical heterogeneity than translational order. Spatial distribution of polar vector $g_i^{\rm max}$ pointing to the direction of maximum $g_i(\theta)$ at {\bf d,} $\phi=0.83$ and {\bf e,} $\phi=0.54$. {\bf f,} Average dot production between particle $i$'s polar vector $g_i^{\rm max}$ and particle $j$'s polar vector $g_j^{\rm max}$ at the distance $r$ to the particle $i$. The polar vectors are correlated within three shells. }
  \label{fig2}
\end{figure*}

Fig.~\ref{fig1}b and Fig.~\ref{fig1}c plot the spatial distribution of  fluctuations of local ADF ($O_{i}$) and that of local RDF ($T_{i}$) for each particle in a colloidal glass with packing fraction $\phi\sim0.85$. To remove short-time fluctuations, $O_{i}$ and $T_{i}$ are extracted from the $g_{i}(r,\theta)$ averaged within beta-relaxation time where the mean square displacement (MSD) reaches plateau~\cite{Yang2, Han1}. Both $O_{i}$ and $T_{i}$ are spatially heterogeneously, and high $O_{i}$ particles are more likely to be clustered than high $T_{i}$ ones. In the following, we compare the spatial heterogeneity of $O_{i}$ (or $T_{i}$) and dynamical heterogeneity as in Fig.~\ref{fig2}c. Dynamical heterogeneity in glasses~\cite{CRR1, DH1, DH2} is characterized by local Debye-Waller factor $\alpha_i = \langle[\vec{r}_i(t)-\vec{r}_i(0)]^2\rangle$, where $\vec{r}_i(t)$ is the position of particle $i$ at time $t$ when the MSD reaches its plateau, and $\left< .\right>$ denotes the time average~\cite{Harrowell1, Tong1}. Comparing Fig.~\ref{fig1}b and Fig.~\ref{fig1}d, we would find that particles locating at high $O_{i}$ regions tend to have higher $\alpha_i$ values. The Spearman's rank correlation coefficient between $O_{i}$ and $\alpha_i$ is as high as $\sim0.8$, significantly higher than that between $T_{i}$ and $\alpha_i$ ($\sim0.5$). The high correlation is robust in samples with different $\phi$ and with $O_{i}$ ($T_{i}$) averaged from a wide range of time $\Delta t$. In Fig.~\ref{fig1}e (Fig.~\ref{fig1}f), a small $\Delta t$ ($\sim 6 s$, when the MSD just reaches its plateau) would saturate the correlation between $O_{i}$ ($T_{i}$) and $\alpha_i$.

In the following, we would focus on how the correlation between the orders we defined and dynamic heterogeneity is developed from liquids. With packing fraction increasing, the dynamics slow down and become more heterogeneous. Fig.~\ref{fig2}a depicts the self-intermediate scattering function $F_{s}(q,t)$ of a series of liquids. $F_{s}(q,t)=\left\langle\sum_{i=1}^N e^{i\bf{q} \cdot[x_{i}(t)-x_{i}(0)]} \right\rangle_{_t}/N$, where $x_{j}(t)$ is the position of particle $i$ at time $t$, $N$ is the number of particles, $\bf{q}$ is the scattering vector determined by the first peak in the structural factor, and $\left< .\right>_{_t}$ denotes a time average. $\alpha$ relaxation time $\tau_{\alpha}$ is then defined as the duration where $F_{s}(q,t)$ decays from $1$ to $1/e$. The dynamic heterogeneity in liquids is characterized by a four-point susceptibility $\chi_4$~\cite{DH3,DH4,DH5,DH6}. $\chi_4$ measures the fluctuations of a two-time self-correlation
function $Q_2$. $Q_2(a,\Delta t)=\frac{1}{N}\sum_{i=1}^N e^{({\frac{-\Delta r^2_i}{2a^2}})}$, where $a$ is the probing spatial scale, $N$ is the number of particles, and $\Delta r^2_i$ is the MSD of particle $i$ within duration $\Delta t$. Then $\chi_4(a,\Delta t)=N(\left\langle Q_2(a,\Delta t)^2 \right\rangle-\left\langle Q_2(a,\Delta t) \right\rangle^2)$, where $\left< .\right>$ stands for a time average. In Fig.~\ref{fig2}b, we plot the dependence of $\chi_4(a,\Delta t)$ on duration $\Delta t$ by preselecting $a$ as the length maximizing $\chi_4$ for each packing. $\chi_4$ reaches its maximum value with $\Delta t$ close to $\alpha$-relaxation time as indicated by the arrows in Fig.~\ref{fig2}b, which is consistent with previous studies of dynamical heterogeneity in different glassy systems~\cite{DH3}. Therefore, the single-particle dynamic in liquids is calculated in $\tau_{\alpha}$, where its dynamic heterogeneous is most pronounced. 

We calculate the correlation between single-particle dynamics and local structural order averaged over different durations $\Delta t_{a}$ in liquids as we do in the glassy solids. The dynamic-structure correlation is maximized in a rather wide $\Delta t_{a}$ range from less than $\frac{1}{10}\tau_{\alpha}$ to $\tau_{\alpha}$~\cite{suppl.}. The maximum correlation for each packing is summarized in Fig.~\ref{fig2}c. With packing fraction increasing, $C^m_{\alpha_i,O_i}$ increases sharply from $0$ to $\sim0.7$, while $C^m_{\alpha_i,T_i}$ changes relatively slowly from $0.1$ to $\sim0.4$. There is a crossover at $\phi_c\sim0.7$ after which $O_i$ becomes better correlated with dynamical heterogeneity than $T_i$. Independent experiments with the same particles suggest that the crossover $\phi_c\sim0.7$ is also the packing fractions where glassy dynamics and dynamic heterogeneity emerge~\cite{Yang3}. After the crossover, orientational order, rather than the translational order, dominates the glassy dynamics, distinguishes glassy liquids from normal liquids. Note that the $C^m_{\alpha_i,T_i}$  in liquids with packing fraction less than $\phi_c\sim0.7$ is $0.1-0.3$. This nontrivial correlation between dynamics and the fluctuations of distribution functions suggests the translational symmetry breaks down even in simple liquids, and could account for the validity of using RDF to predict the macroscopic properties in simple liquids such as diffusion and energy from liquid theories~\cite{liquid1,liquid2,Ghosh1,Wallace1,Mittal1}.

Since the crystalline bond orientational order show neglectable changes with packing increasing for a binary glass with strong geometric frustration, the build-up of $C^m_{\alpha_i,O_i}$ suggests an emergence of amorphous order during glass transition. The specific form of the amorphous order is of interest. Since orientational distribution fluctuations suggests that there are some preferred directions, we plot the spatial distribution of polar vectors pointing to the maximum $g_i(\theta)$ for each particle. It seems that there are some correlated domains for vectors in supercooled liquids with $\phi=0.83$ (In Fig.~\ref{fig2}d), compared to the random distributions of vectors in normal loose liquids (Fig.~\ref{fig2}e). The correlation function among vectors are defined as the sum of the dot-product between the polar vector of reference particle $i$ and those of the neighboring particles at a given distance $r$. While the correlation is trivial at all distances for loose liquids, there is a significant peak for the liquids approaching jamming ($\phi_j\sim0.85$) as in Fig.~\ref{fig2}f. The correlation peak also emerges close to the crossover packing $\phi_c$ where $C^m_{\alpha_i,O_i}$ surpasses $C^m_{\alpha_i,T_i}$. The correlation length of orientational order is about three shells of neighboring particles. More detailed features of the amorphous order remain to be revealed.

In conclusion, the orientational order defined from a particle's angular distribution fluctuations are demonstrated to be better correlated with dynamic heterogeneity than the translational order extracted from radial distribution functions. During glass transition, particles tune their orientational distributions and form orientational orders besides the simple increase of density. Most popular liquid state theories such as the density functional theory~\cite{DFT1, DFT2} and mode-coupling theory ~\cite{MCT1,MCT2} only consider the scalar density field (the translational order) and two-body effects. Our experimental results reveal that the non-local orientational order plays even more important roles in strongly disordered systems than translational order. The orientational order emphasizes many-body effects in complex liquids and may call for a rethink for the structure in dense liquids or glassy systems. Particle distribution fluctuations are irrelevant to inter-particle interactions and unlimited to specific polydisperse systems, and thus could be borrowed to other strongly disordered complex systems such as multi-component metallic glasses~\cite{Metallic} and vibrated granular systems~\cite{Paddy2}.

We thank Rui Liu, Mingcheng Yang, and Chenhong Wang for helpful discussions. This work was supported by the MOST 973 Program (No. 2015CB856800). K. C. also acknowledges the support from the NSFC (No. 11474327).


\begin{thebibliography}{47}

\item[$^*$]yxn@iphy.ac.cn
\item[$^*$]kechen@iphy.ac.cn
\item[$^*$]whw@iphy.ac.cn\\



\bibitem{Grfail1}A. van Blaaderen, P. Wiltzius, Science {\bf 270}, 1177 (1995).
\bibitem{Grfail2}A. Cavagna, Phys. Rep. {\bf 476}, 51 (2009).

\bibitem{DH1} L. Berthier and G. Biroli, Rev. Mod. Phys.{\bf 83}, 587 (2011).
\bibitem{DH2} L. Berthier, G. Biroli, J.-P. Bouchaud, L. Cipeletti, and W. van Saarloos, \emph{Dynamical Heterogeneities in Glasses,
Colloids, and Granular Media} (Oxford University Press,Oxford, England, 2011).

\bibitem{DH3} N. Lačević, F. W. Starr, T. B. Schrøder, and S. C. Glotzer, J. Chem. Phys. {\bf 119}, 7372 (2003).
\bibitem{DH4} Z. Zhang, P. J. Yunker, P. Habdas, and A. G. Yodh, Phys. Rev. Lett. {\bf 107},208303 (2011).

\bibitem{DH5} A. S. Keys et al., Nature Phys. {\bf 3}, 260 (2007).

\bibitem{DH6} O. Dauchot, G. Marty, and G. Biroli, Phys. Rev. Lett. {\bf 95}, 265701 (2005).
\bibitem{DH7} Ediger, M. D. Annu. Rev. Phys. Chem. 51,99 (2000). 
\bibitem{DH8} W. Kob, C. Donati, S. J. Plimpton, P. H. Poole, and S. C. Glotzer, Phys. Rev. Lett. {\bf 79}, 2827 (1997).
\bibitem{DH9} W. K. Kegel, and Alfons van Blaaderen, Science {\bf 287}, 290 (2000).
\bibitem{DH10} E. R. Weeks \emph{et al.} Science {\bf 287}, 627 (2000).
\bibitem{DH11} L. Berthier et al., Science {\bf 310}, 1797 (2005).

\bibitem{Tanaka1}H. Tanaka, T. Kawasaki, H. Shintani, and K. Watanabe, Nature Mater. {\bf 9}, 324 (2010).
\bibitem{sym_2}H. W. Sheng, W. K. Luo, F. M. Alamgir, J. M. Bai, and E. Ma, Nature {\bf 439}, 419 (2006).
\bibitem{sym_3}K. F. Kelton \emph{et al.} Phys. Rev. Lett. {\bf 90}, 195504 (2003).
\bibitem{sym_4}Y.-C. Hu, F.-X. Li, M.-Z. Li, H.-Y. Bai, and W.-H. Wang, Nat. Commun. {\bf 6}, 8310 (2015).
\bibitem{sym_5}R. Milkus and A. Zaccone, Phys. Rev. B {\bf 93},094204 (2016).

\bibitem{PTS1} G. Biroli, J.-P. Bouchaud, A. Cavagna, T. S. Grigera, and P. Verrocchio, Nature Phys. {\bf 4}, 771 (2008).
\bibitem{PTS2} K. Hima Nagamanasa, S. Gokhale, A. K. Sood, and R. Ganapathy, Nature Phys. {\bf 11}, 403 (2015).
\bibitem{PTS3} B. Zhang and X. Cheng, Phys. Rev. Lett. {\bf 116}, 098302 (2016).

\bibitem{PTS4} G. M. Hocky, T. E. Markland, and D. R. Reichman, Phys. Rev. Lett. {\bf 108}, 225506 (2012).
\bibitem{PTS5} J. Russo and H. Tanaka, Proc. Natl. Acad.
Sci. U.S.A. {\bf 112}, 6920 (2015).
\bibitem{Paddy1} C. P. Royall and S. R. Williams, Physics Reports {\bf 1}, 560 (2015).


\bibitem{Andrea1} E. D. Cubuk, S. S. Schoenholz, E. Kaxiras, and A. J. Liu, J. Phys. Chem. B {\bf 120}, 6139 (2016).
\bibitem{Andrea2} E. D. Cubuk \emph{et al.} Phys. Rev. Lett. {\bf 114}, 108001 (2015).
\bibitem{Andrea3} S. S. Schoenholz, E. D. Cubuk, D. M. Sussman, E. Kaxiras, and A. J. Liu, Nature Phys. {\bf 12}, 469 (2016).
\bibitem{Andrea4} Cubuk et al., Science {\bf 358},  1033 (2017).


\bibitem{Yang2} X. Yang, R. Liu, M. Yang, W.-H. Wang, and K. Chen, Phys. Rev. Lett. {\bf 116}, 238003 (2016).
\bibitem{Tanaka2} T. Kawasaki, T. Araki, and H. Tanaka, Phys. Rev. Lett. {\bf 99}, 215701 (2007).
\bibitem{Han1}Z. Zheng \emph{et al.} Nat. Commun. {\bf 5}, 3829 (2014).
\bibitem{Tanaka3}H. Tanaka, Eur. Phys. J. E {\bf 35}, 113 (2012).

\bibitem{Yunker1}P. J. Yunker et al. Rep. Prog. Phys. {\bf 77}, 056601 (2014).
\bibitem{Chen2010}K. Chen \emph{et al.} Phys. Rev. Lett. {\bf 105}, 025501 (2010).
\bibitem{Chen2011}K. Chen \emph{et al.} Phys. Rev. Lett. {\bf 107}, 108301 (2011).

\bibitem{Grier1}J. C. Crocker, and D. G. Grier, J. Colloid Interface Sci. {\bf 179}, 298 (1996).

\bibitem{suppl.}
See EPAPS Document No. XXX for a discussion of additional experimental details.
For more information on EPAPS, see http://www.aip.org/pubservs/epaps.html


\bibitem{CRR1} G. Adam and J. H. Gibbs, J. Chem. Phys. {\bf 43}, 139 (1965).



\bibitem{Harrowell1}A. Widmer-Cooper and P. Harrowell, Phys. Rev. Lett. {\bf 96}, 185701 (2006).
\bibitem{Tong1}H. Tong and N. Xu, Phys. Rev. E {\bf 90}, 010401 (2014).



\bibitem{Yang3} X. Yang, T. Hua W.-H, Wang, K. Chen, http://arxiv.org/abs/1710.08154.





\bibitem{liquid1} Jean-Pierre Hansen, Ian R. McDonald \emph{Theory of Simple Liquids with Applications to Soft Matter (Fourth Edition)} (Academic Press, Elsevier Ltd., 2013).

\bibitem{liquid2} Born, M. and Green M.S., \emph{A General Kinetic Theory of Liquids} (Cambridge University Press, Cambridge, 1949).
\bibitem{Ghosh1}A. Samanta, Sk. Musharaf Ali, and S. K. Ghosh, Phys. Rev. Lett. {\bf 92}, 145901 (2004).
\bibitem{Wallace1}D. C. Wallace, J. Chem. Phys. {\bf 87}, 2282 (1987).
\bibitem{Mittal1}J. Mittal, J. R. Errington, and T.M. Truskett, J. Phys. Chem. B  {\bf 110}, 18147 (2006).

\bibitem{DFT1} R. Car, and M. Parrinello, Phys. Rev. Lett. {\bf 55}, 2471 (1985).
\bibitem{DFT2} J. L. Barrat, M. Baus, and J. P. Hansen, Phys. Rev. Lett. {\bf 56}, 1063 (1986).

\bibitem{MCT1} W. Van Megen, and S. M. Underwood, Phys. Rev. E {\bf 49}, 4206 (1994).
\bibitem{MCT2} W. Kob, and H. C. Andersen, Phys. Rev. E {\bf 51}, 4626 (1995).

\bibitem{Metallic}Y. H. Liu \emph{et al.}, Science {\bf 315}, 1385 (2007).

\bibitem{Paddy2}R. Patrick \emph{et al.}, Nature Mater. {\bf 4}, 121 (2005).

\end{thebibliography}
\end{document}